\documentclass{article}
\usepackage{amsmath}
\usepackage{graphicx} 
\usepackage[leftcaption]{sidecap} 
\usepackage{multicol}
\usepackage{caption}
\usepackage{float}

\usepackage[backend=biber, style=phys, url=true]{biblatex}
\usepackage[colorlinks=true, linkcolor=blue, citecolor=blue, urlcolor=blue]{hyperref}

\title{Studying the onset of deconfinement at NA61/SHINE}

\author{Syed Uzair Ahmed Shah\textsuperscript{1} for the NA61/SHINE Collaboration\\
[11pt]
\textsuperscript{1}Jan Kochanowski University, Kielce, Poland
}

\date{}

\begin{document}

\maketitle

\textbf{Abstract:} The study of particle production in high-energy collisions of heavy ions offers a unique opportunity to explore the phase transition of strongly interacting matter. The NA61/SHINE experiment, a fixed-target setup located in the CERN SPS North Area, plays a crucial role in this investigation. To perform a two-dimensional scan of the phase diagram of strongly interacting matter, the experiment varies both the beam momentum (ranging from 13 to 150/158 $\mathrm{GeV}/c$ per nucleon) and the size of the colliding ions (\textit{p+p}, \textit{p}+Pb, Be+Be, Ar+Sc, Xe+La, Pb+Pb). This article presents a study of the properties of the onset of deconfinement by measuring the $K^{+}/\pi^{+}$ ratio, which is proportional to the strangeness to entropy ratio of the produced system. New results from central Pb+Pb collisions at $30A$ $\mathrm{GeV}/c$, focusing on the spectra of charged kaons and pions, confirm the presence of the \textit{horn} structure in the energy dependence of the $K^{+}/\pi^{+}$ ratio, consistent with earlier observations by NA49 and STAR, which could be interpreted as a hint of the onset of deconfinement at middle SPS energies. 

\section*{Introduction}
The study of the phase transition from confined hadronic matter to a deconfined state of quarks and gluons, the quark-gluon plasma (QGP) \cite{Shuryak:1978ij}, remains a central topic in high-energy nuclear physics. The NA61/SHINE experiment at the CERN Super Proton Synchrotron (SPS) \cite{Abgrall:2014fa} is dedicated to explore this transition through a systematic study of hadron production across varying collision energies and system sizes. This analysis is particularly focused on identifying signals of the onset of deconfinement.

\begin{figure}[H]
    \centering
    \includegraphics[width=0.94\linewidth]{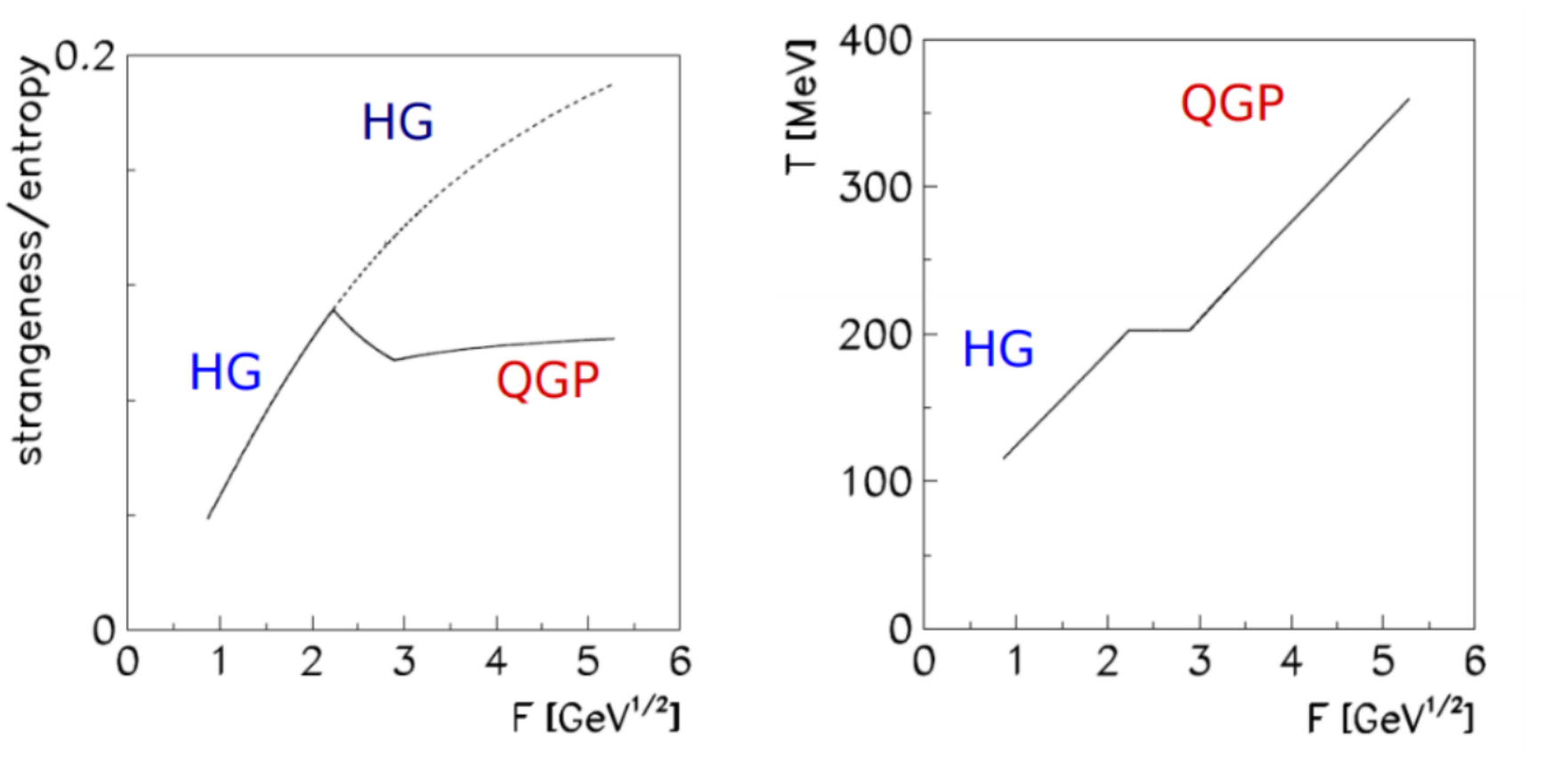}
    \vspace{-0.4cm}
    \caption{Sharp peak in the strangeness to entropy ratio (\textit{left}) and plateau in system temperature (\textit{right}) at the phase transition, shown as functions of the Fermi variable $F\approx (s_{NN})^{1/4}$, where $\sqrt{s_{NN}}$ is the collision center-of-mass energy per nucleon pair.}
    \label{fig:SMES}
\end{figure}

The study is based on the prediction of the Statistical Model of the Early Stage \cite{Gazdzicki:1998vd}, and the results from the NA49 experiment, which observed significant changes in hadron production properties in central Pb+Pb collisions around a beam momentum of $30A$ $\mathrm{GeV}/c$~\cite{NA49:2007stj}. These features include a nonmonotonic energy dependence of the $K^{+}/\pi^{+}$ ratio (the so-called \textit{horn}) and a plateau in the inverse slope parameter of the $K^{+}$ and $K^{-}$ transverse momentum/mass spectra (the \textit{step}) which were interpreted as signatures of the onset of the QGP phase following predictions presented in Fig. \ref{fig:SMES}. To study these signatures, NA61/SHINE performs a two-dimensional scan by varying both the beam momentum from 13 to 150/158 $\mathrm{GeV}/c$ per nucleon and the size of the colliding nuclei \textit{p+p}, \textit{p}+Pb, Be+Be, Ar+Sc, Xe+La, Pb+Pb. Increasing collision energy tends to increase the temperature and decrease the baryon chemical potential at freeze-out, while increasing system size generally results in lower freeze-out temperatures \cite{PhysRevC.73.044905}. It enables the understanding of the system size dependence of the \textit{horn} and \textit{step} signals.

The NA61/SHINE experiment benefits from a fixed target design, which offers excellent acceptance in the forward rapidity region and at very low transverse momentum ($p_{\text{T}} = 0$) \cite{Abgrall:2014fa}, thereby enabling precise measurements of identified hadron spectra. Its primary detection system includes large-volume Time Projection Chambers (TPCs), time-of-flight (ToF) walls, and a high-resolution Projectile Spectator Detector (PSD) used for centrality determination \cite{Golubeva:2012zz}.

In the present study, new measurements are presented from central Pb+Pb collisions at $30A$ $\mathrm{GeV}/c$. These results include $p_{\text{T}}$ spectra and rapidity distributions for $K^{+}$, $K^{-}$, and $\pi^{-}$. The extracted value of the $K^{+}/\pi^{+}$ ratio is consistent with the \textit{horn} structure previously reported.

\section*{Particle identification}
In the NA61/SHINE experiment, particle identification is achieved through two complementary techniques: measurement of the ionization energy loss ($\mathrm{d}E/\mathrm{d}x$) in TPCs and the time of flight of the particle in ToF detectors~\cite{NA61SHINE:2023epu}. Together, these methods provide reliable identification of $\pi^{+}, \pi^{-}, K^{+}, K^{-}, p, \bar{p}$ and other charged hadrons across a wide range of momenta and rapidities. At high momenta, the ionization energy loss measured in the TPC gas increases gradually, and particles of different masses can be distinguished based on their characteristic $\mathrm{d}E/\mathrm{d}x$ behavior. At lower momenta, $\mathrm{d}E/\mathrm{d}x$ curves for various particle

\begin{figure}[H]
    \centering
    \vspace{-0.3cm}
    \includegraphics[width=1.0\textwidth]{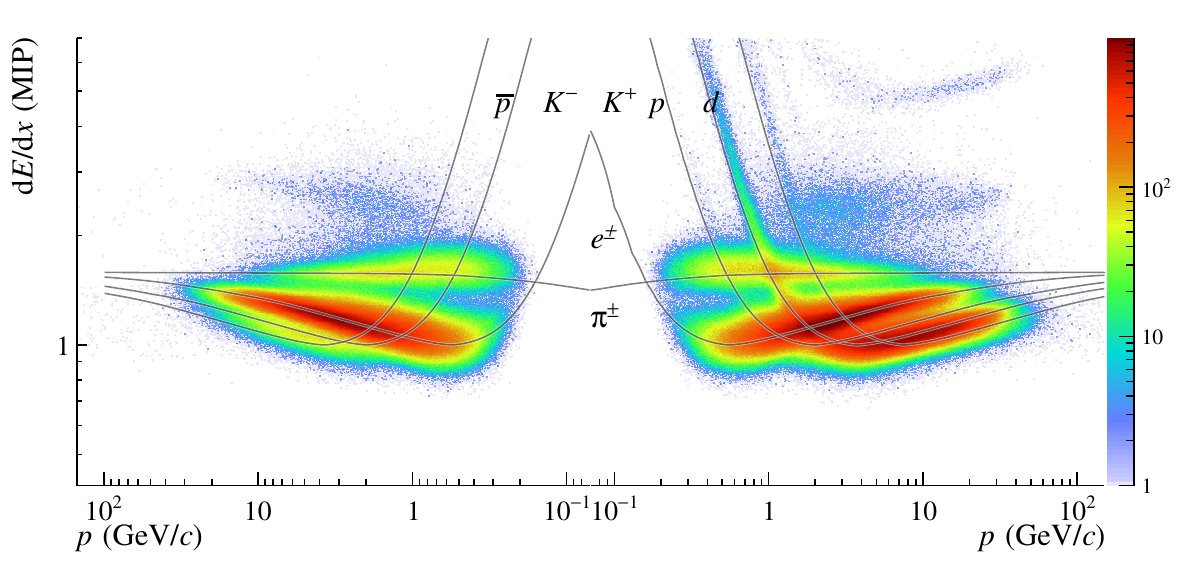}
    \vspace{-0.7cm}
    \caption{The ionization energy loss of particles in TPCs versus total momentum (in laboratory frame) together with Bethe-Bloch curves. Plot for 0--7.2\% central Pb+Pb collisions at $30A$ $\mathrm{GeV}/c$. }
    \label{fig:dedxVsP}
\end{figure}
\begin{center}
\begin{minipage}{0.48\textwidth} 
    \vspace{-0.2cm}
    species begin to overlap, as can be seen in Fig. \ref{fig:dedxVsP}. To resolve this, ToF measurements are used in conjunction with $\mathrm{d}E/\mathrm{d}x$ to improve identification, especially for kaons near mid-rapidity. To extract particle yields, the $\mathrm{d}E/\mathrm{d}x$ distributions are analyzed in small binned intervals of total momentum (\textit{p}) and transverse momentum. Figure~\ref{fig:dedx_fit} presents an example of $\mathrm{d}E/\mathrm{d}x$ in one such bin fitted with a sum of asymmetric Gaussian functions~\cite{vanLeeuwen:2003it}, each representing a different particle type ($d$, $p$, $K^{+}$, $\pi^{+}$, $e^{+}$) and weighted by track length (number of clusters $l$). The fit function is described by the formula:
\end{minipage}
\hfill
\begin{minipage}{0.48\textwidth}
    \centering
    \vspace{-0.5cm}
    \includegraphics[width=1.\textwidth]{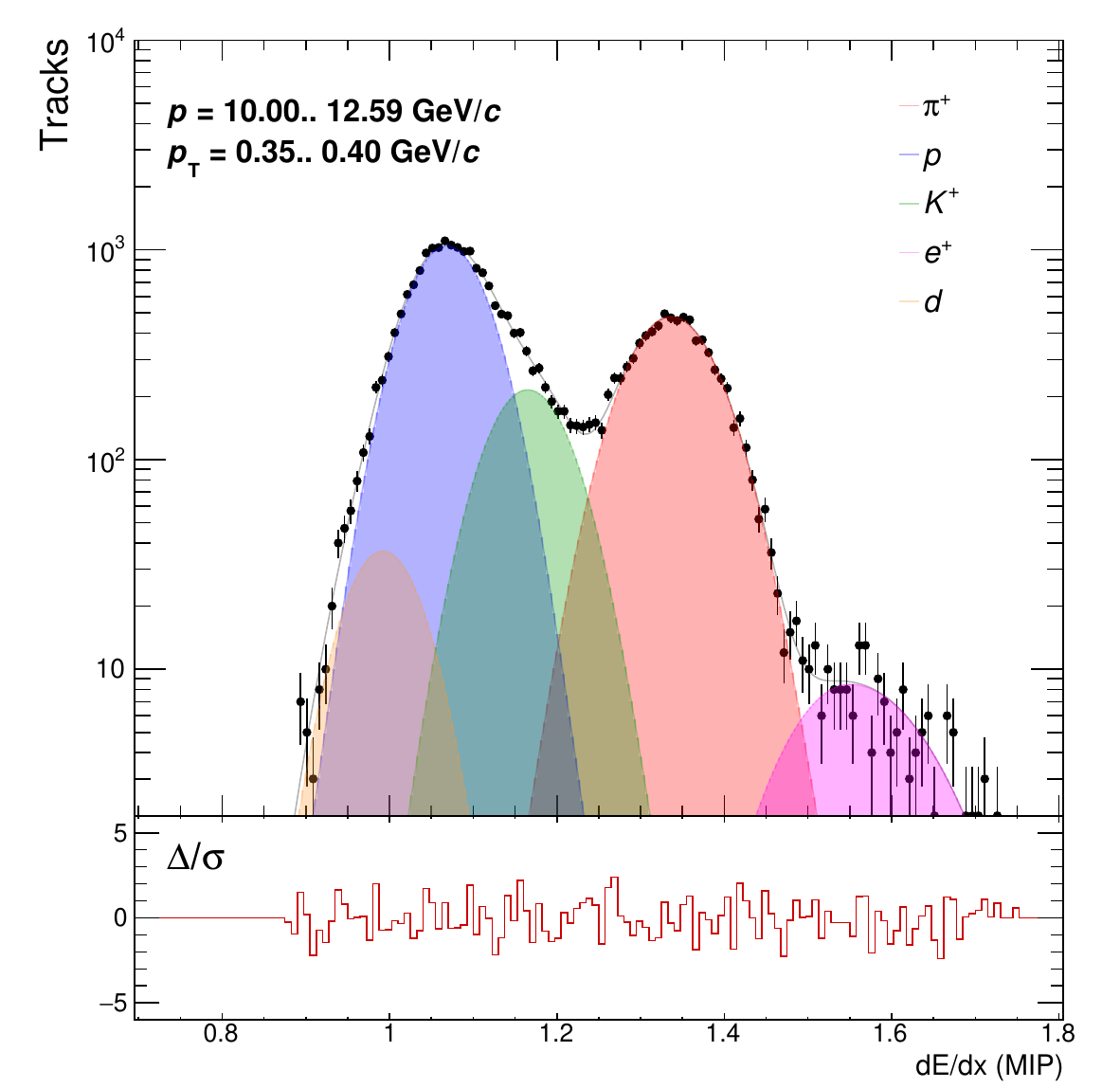}
    \vspace{-0.7cm} 
    \captionof{figure}{The example of $\mathrm{d}E/\mathrm{d}x$ distribution measured in 0--7.2\% central Pb+Pb collisions at $30A$ $\mathrm{GeV}/c$.}
    \label{fig:dedx_fit}
\end{minipage}
\end{center}
\begin{equation} \label{eq:1}
   f(x) = \sum_{i=\pi,K,e,p,d} N_i \frac{1}{\sum_l n_l} \sum_l \frac{n_l}{\sigma_{i,l} \sqrt {2\pi}} \exp\Biggl[-\frac{1}{2} \Bigl( \frac{x-x_i^{\prime}}{(1 \pm \delta)\sigma_{i,l}}\Bigr)^2 
  \Biggr],
\end{equation}
where $N_i$ is the amplitude of the particles $i$ and $n_l$ is the number of tracks with $l$. The width $\sigma_{i,l}$ depends on the number of clusters and particle type, scaling with $\left(\frac{x_i}{x_\pi}\right)^\alpha$, where $\alpha = 0.625$. The peak position $x_i^{\prime} (= x_i - \frac{2}{\sqrt{2\pi}} \delta\sigma_{i,l})$ is shifted for $x_{i}$ to account for the asymmetry. The shift depends on a parameter $\delta = \delta_0/l$, where $\delta_0$ describes the asymmetry in distribution.

Additionally, a probabilistic approach is used to assign particle identities~\cite{Gazdzicki:2011xz}. For each track, the likelihood of it being a particular species is calculated based on its $p$, $p_{\mathrm{T}}$ and measured $\mathrm{d}E/\mathrm{d}x$:
\begin{equation} \label{eq:2}
    P_i^{p,p_\mathrm{T}}\text{\small $(dE/dx)$} = \frac{\rho_i^{p,p_\mathrm{T}} \text{\small $(\mathrm{d}E/\mathrm{d}x)$}}{\sum_{i=\pi^{\pm}, K^{\pm}, p, \bar{p}, e^{\pm}, d} \rho_i^{p,p_\mathrm{T}}\text{\small $(\mathrm{d}E/\mathrm{d}x)$}}.
\end{equation}
Here, $\rho_i^{p, p_T}(\mathrm{d}E/\mathrm{d}x)$ is the fitted energy loss distribution for particle type $i$ in the bin. Using these probabilities, the raw yield $n_i$ of particles of type $i$ in a bin can be determined by summing over all tracks, as in equation:
\begin{equation} \label{eq:3}
    n_i =\frac{\sum_{j=1}^m P_i}{N_{events}},
\end{equation}
where $m$ is the total number of tracks, $i$ runs over the relevant species ($\pi^{\pm}, K^{\pm}, p,$ $\bar{p}, e^{\pm}, d$), and $N_{events}$ is the number of accepted events. This probability is derived from parametrization fits of Eq.~\ref{eq:1} to the experimental data in each ($p, p_\mathrm{T}$) bin, allowing for statistical separation of species even in overlapping regions.

\section*{Results}
In this section, the preliminary results on $ K^{+}$, $K^{-}$ and $\pi^{-}$ production in central Pb+Pb collisions at beam momentum 30$A$ $\mathrm{GeV}/c$ are presented. The spectra obtained using $\mathrm{d}E/\mathrm{d}x$ method, are corrected for detector acceptance, tracking and reconstruction efficiency, as well as contributions from weak decays and secondary interactions. These corrections were performed using Monte Carlo simulations based on the EPOS1.99~\cite{Werner:2008zza} coupled with GEANT4 \cite{Agostinelli:2002hh}.

\begin{figure}[H]
    \centering
    \vspace{-1cm}
    \includegraphics[width=1.\textwidth]{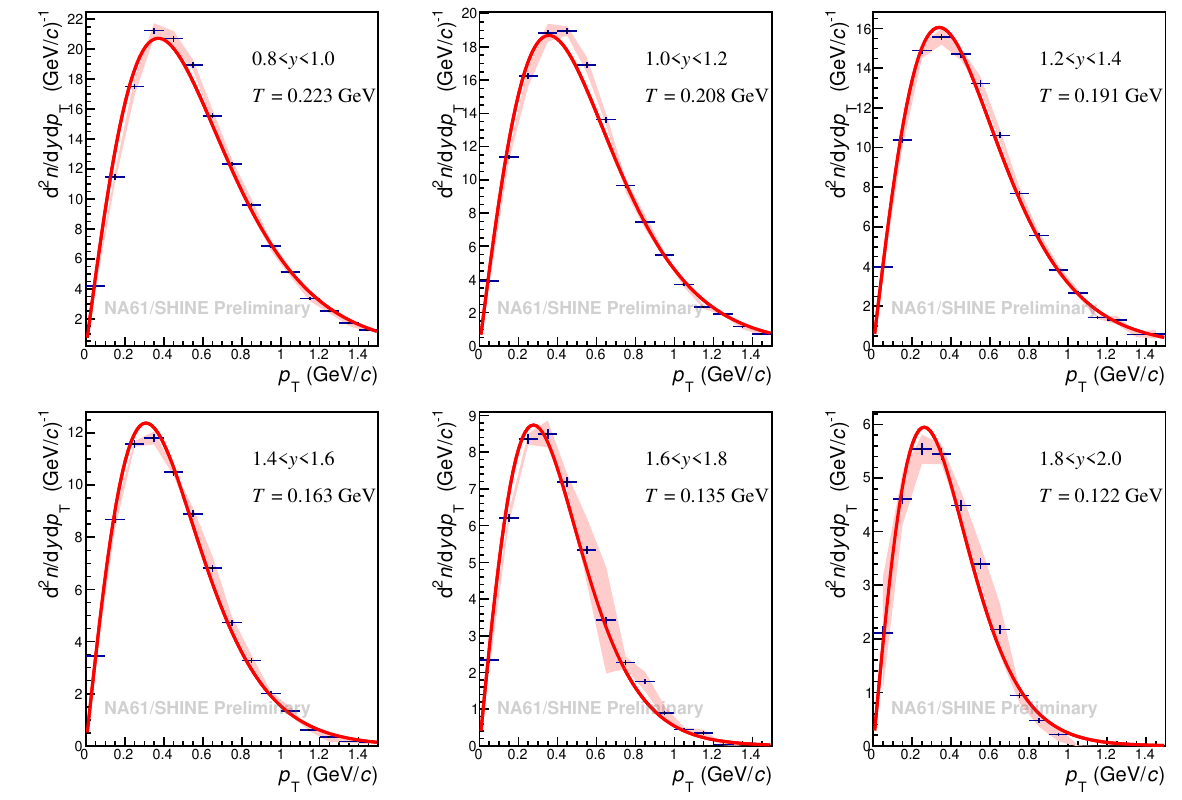}
    \vspace{-0.7cm}
    \caption{The $\mathrm{d}^2n / \mathrm{d}y\mathrm{d}p_\mathrm{T}$ spectra of $K^+$ mesons measured in 0--7.2\% central Pb+Pb collisions at $30A$ $\mathrm{GeV}/c$. The rapidity ($y$) values are given in the collision center-of-mass reference system. Vertical lines represent statistical uncertainties, while systematic uncertainties are indicated by red bands.}
    \label{fig:pTpos}
\end{figure}

\begin{figure}[H]
    \centering
    \vspace{-0.6cm}
    \includegraphics[width=1.\textwidth]{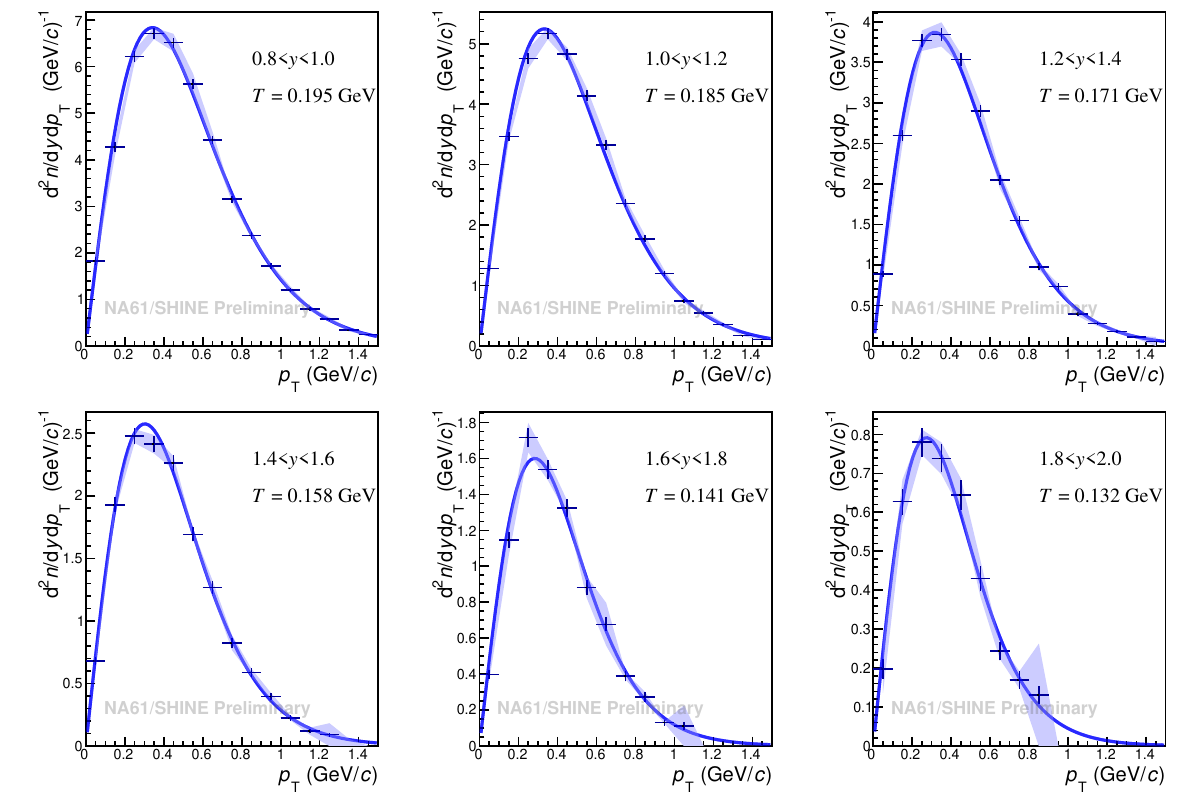}
    \vspace{-0.7cm}
    \caption{The $\mathrm{d}^2n / \mathrm{d}y\mathrm{d}p_\mathrm{T}$ spectra of $K^-$ mesons measured in 0--7.2\% central Pb+Pb collisions at $30A$ $\mathrm{GeV}/c$. The rapidity ($y$) values are given in the collision center-of-mass reference system. Vertical lines represent statistical uncertainties, while blue bands indicate systematic uncertainties.}
    \label{fig:pTneg}
\end{figure}

The double-differential spectra of identified $K^+$ and $K^-$ mesons produced in the 7.2\% most central Pb+Pb collisions are shown in Fig.~\ref{fig:pTpos} and \ref{fig:pTneg}, respectively.  The $p_{\text{T}}$ spectra are fitted with the function: 
\begin{equation} \label{eq:4}
    f(p_{\mathrm{T}})= A \cdot p_{\text{T}} \cdot e^{-\frac{m_{\text{T}} - m_K}{{T}}},
\end{equation}
where $m_K$ and $m_\mathrm {T}$ represent the mass and transverse mass of the kaon, respectively, and $T$ is the inverse slope parameter.

\begin{figure}[H]
    \centering
    \includegraphics[width=0.49\textwidth]{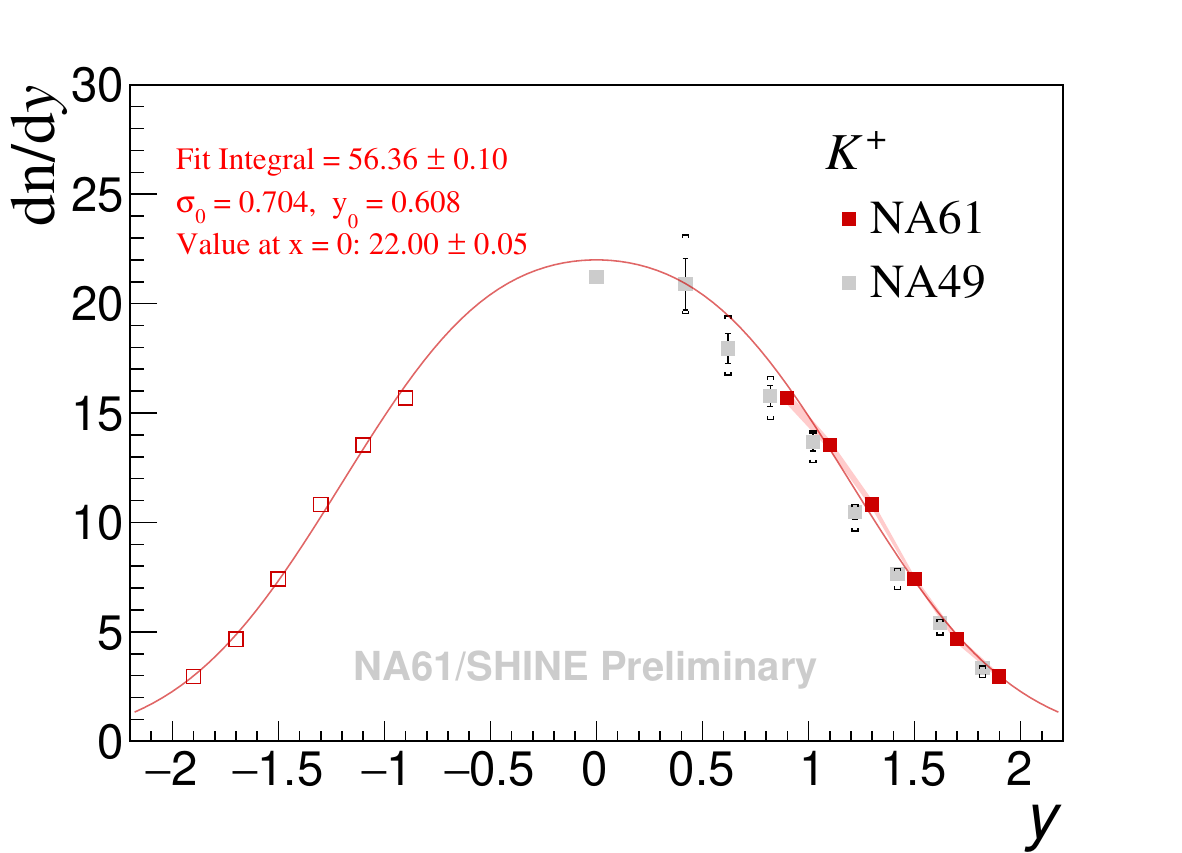}
    \includegraphics[width=0.49\textwidth]{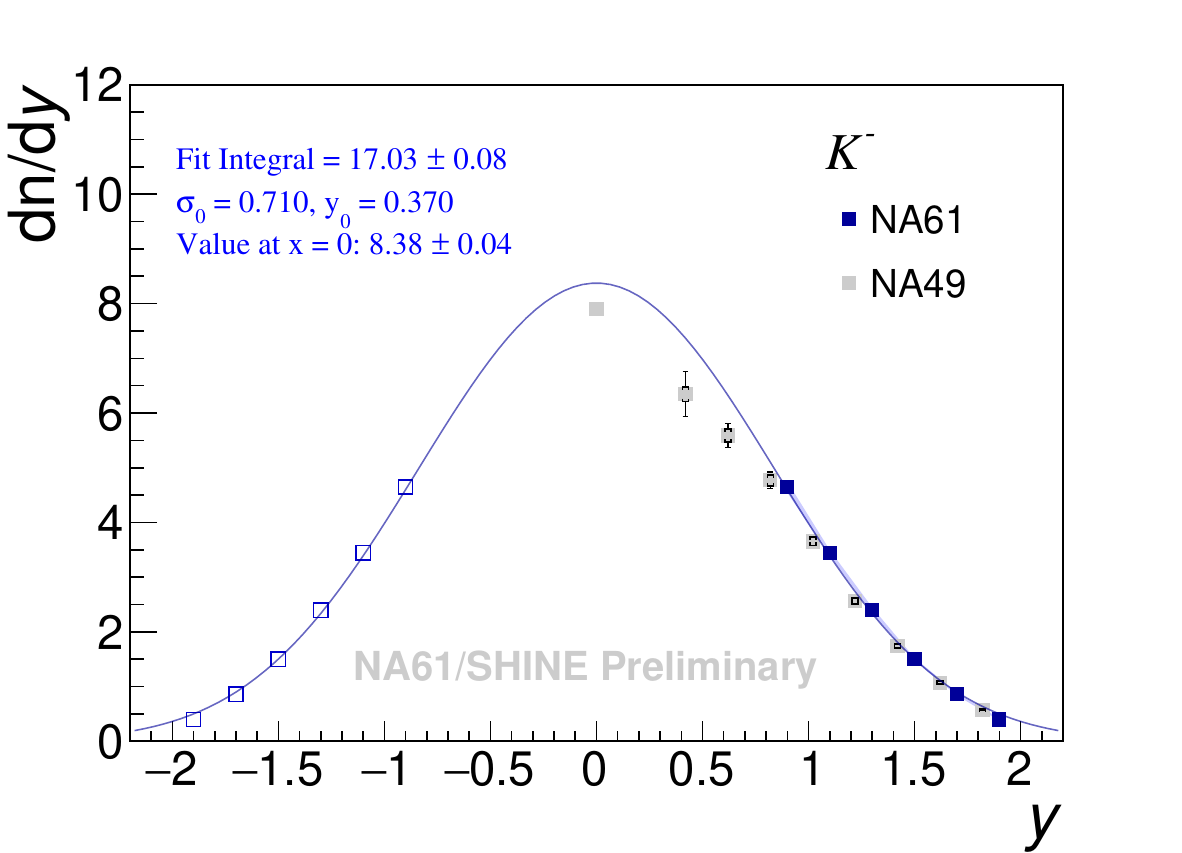}
    \vspace{-0.2cm}
    \caption{The rapidity spectra of $K^+$ and $K^-$ mesons measured in 0--7.2\% central Pb+Pb collisions at $30A$ $\mathrm{GeV}/c$ presented with statistical (vertical bars) and systematic (color bands or brackets) uncertainties. Full symbols indicate the measured data, whereas open symbols correspond to their reflection around mid-rapidity. NA49 points are taken from Ref.~\cite{NA49:2007stj}.}
    \label{fig:kaonY}
\end{figure}

The $\mathrm{d}n/\mathrm{d}y$ distributions of identified $K^+$ and $K^-$ mesons are shown in Fig.~\ref{fig:kaonY}, fitted with a double gaussian function:
\begin{equation} \label{eq:5}
    f(y) = \frac{A}{\sigma_{0}\sqrt{2\pi}} \Bigl[ e^{\frac{-(y-y_0)^2}{2\sigma_{0}^2}}+e^{\frac{-(y+y_{0})^2}{2\sigma_{0}^2}}\Bigr],
\end{equation}
where the fit parameters $\sigma_{0}$ and $y_0$ are taken from NA49 \cite{NA49:2007stj}.


\begin{figure}[H]
    \includegraphics[width=\textwidth]{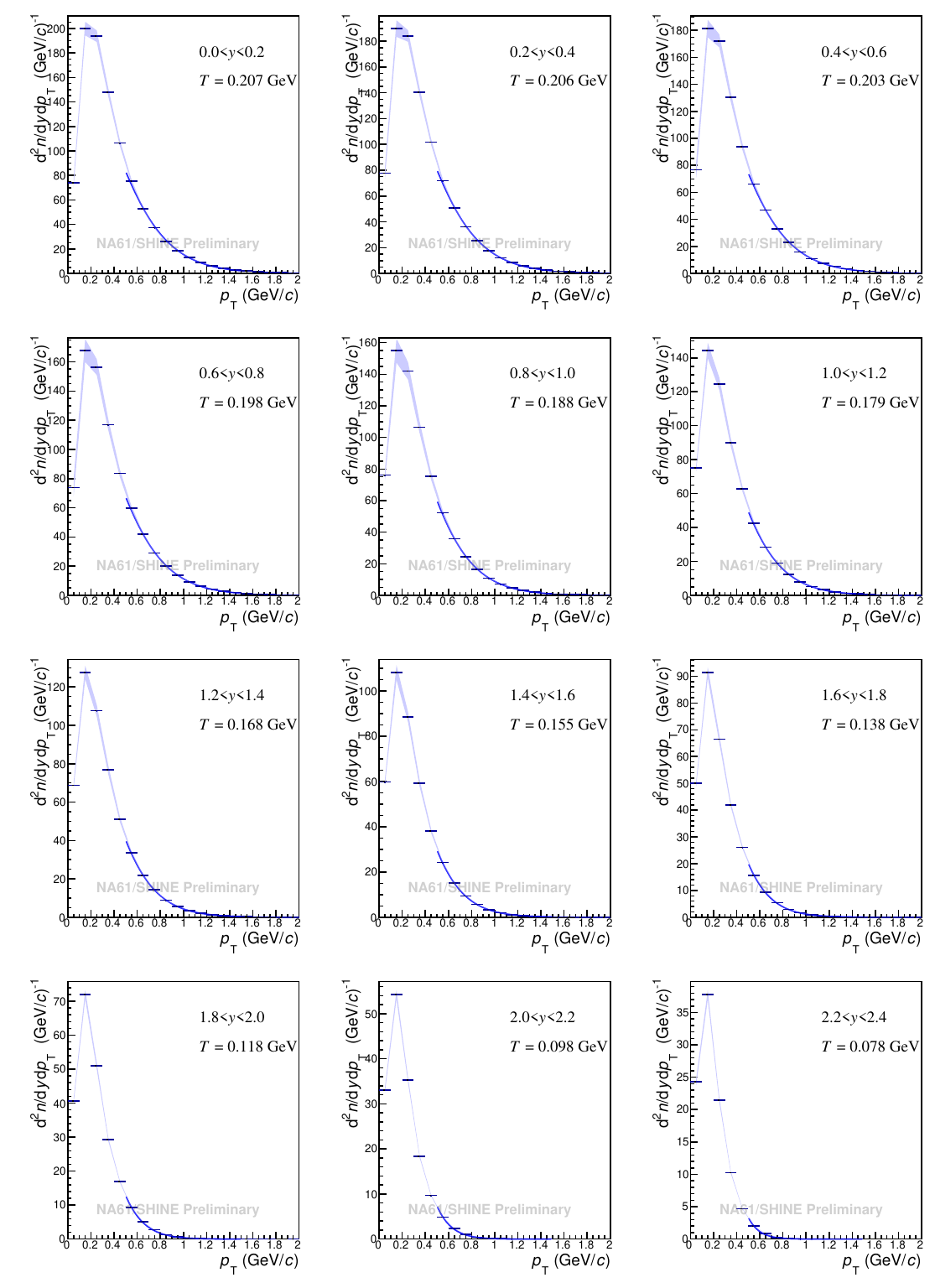}
    \vspace{-0.8cm}
    \caption{The $\mathrm{d}^2n / \mathrm{d}y\mathrm{d}p_\mathrm{T}$ spectra of $\pi^-$ mesons measured in 0--7.2\% central Pb+Pb collisions at $30A$ $\mathrm{GeV}/c$. The rapidity ($y$) values are given in the collision center-of-mass reference system. Vertical lines represent statistical uncertainties, while blue bands indicate systematic uncertainties.}
    \label{fig:pionPT}
\end{figure}

The spectra of $\pi^-$ are obtained using the $h^-$ analysis technique. This method utilizes the fact that most of the negatively charged particles produced in heavy-ion collisions are $\pi^-$ mesons (90\% approximately) \cite{NA61SHINE:2013tiv}. The smaller contributions from $K^-$ and a negligible number of $\bar{p}$ are subtracted using EPOS1.99 simulations. This approach allows for large-acceptance measurements of pion spectra, particularly valuable for determining total yields and comparing production rates across different particle species. Figures~\ref{fig:pionPT} and \ref{fig:pionY} present the $\mathrm{d}^2n/\mathrm{d}y\mathrm{d}p_\mathrm{T}$ and $\mathrm{d}n/\mathrm{d}y$ spectra of $\pi^-$, respectively.

\begin{center}
\begin{minipage}{0.45\textwidth} 
    \vspace{0.2cm}
    \captionof{figure}{The rapidity spectra of $\pi^-$ mesons measured in 0--7.2\% central Pb+Pb collisions at $30A$ $\mathrm{GeV}/c$. NA61/SHINE results are preliminary, NA49 points are taken from Ref.~\cite{NA49:2007stj}. Statistical uncertainties are smaller than the marker size. For NA61/SHINE points, the systematic uncertainties are shown as red bands.} 
    \label{fig:pionY}
\end{minipage}
\hfill
\begin{minipage}{0.52\textwidth}
    \centering
    \vspace{-0.2cm}
    \includegraphics[scale=0.33]{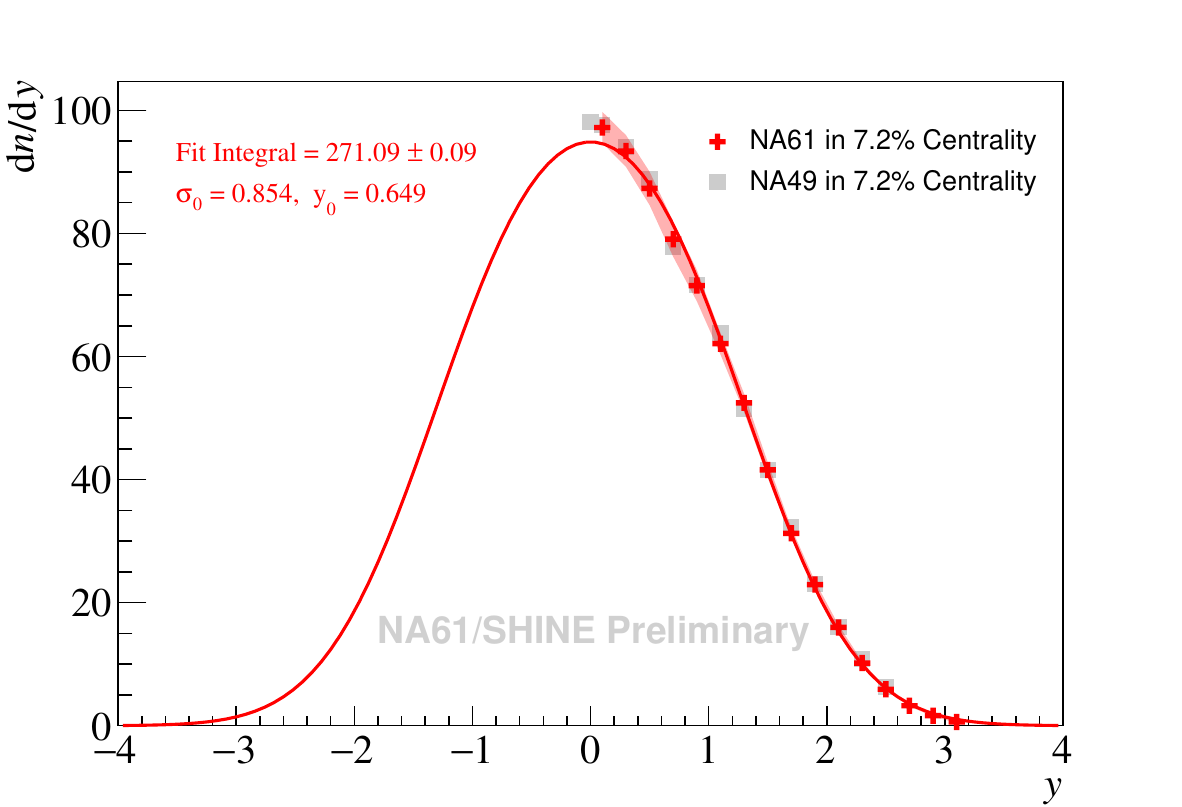}
\end{minipage}
\end{center}

\section*{Onset of Deconfinement}
To calculate the $K^+/\pi^+$ ratio, the total yield of $K^+$ is determined by adding the measured $\mathrm{d}n/\mathrm{d}y$ values and the contribution from the unmeasured rapidity regions using an extrapolated fit. The mean multiplicity of \( \pi^+ \) is estimated by scaling the measured $\pi^-$ yield with the $\pi^+/\pi^-$ ratio obtained by the NA49 experiment \cite{NA49:2007stj}.

\begin{figure}[h]
    \includegraphics[scale=0.3]{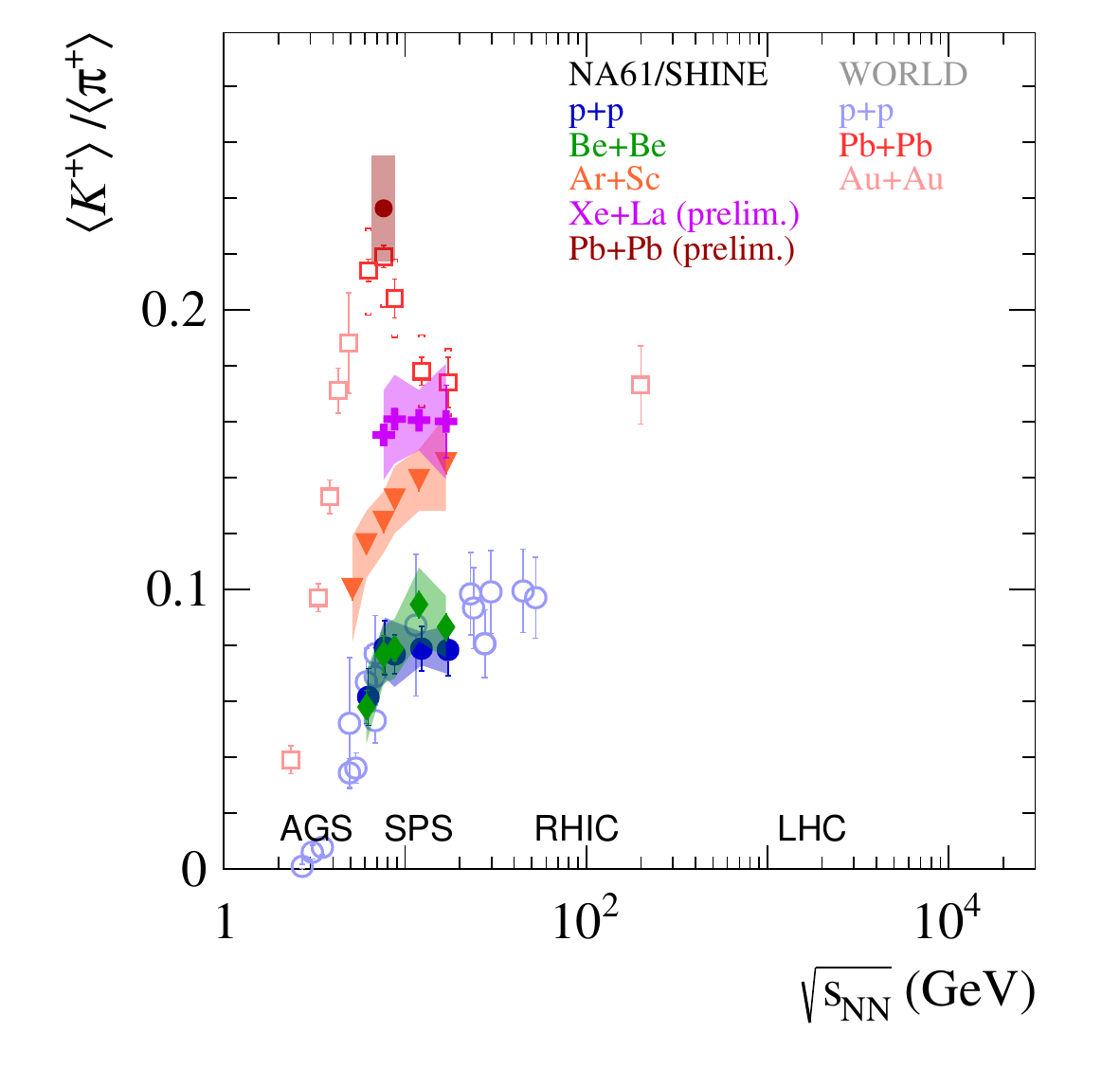}
    \includegraphics[scale=0.3]{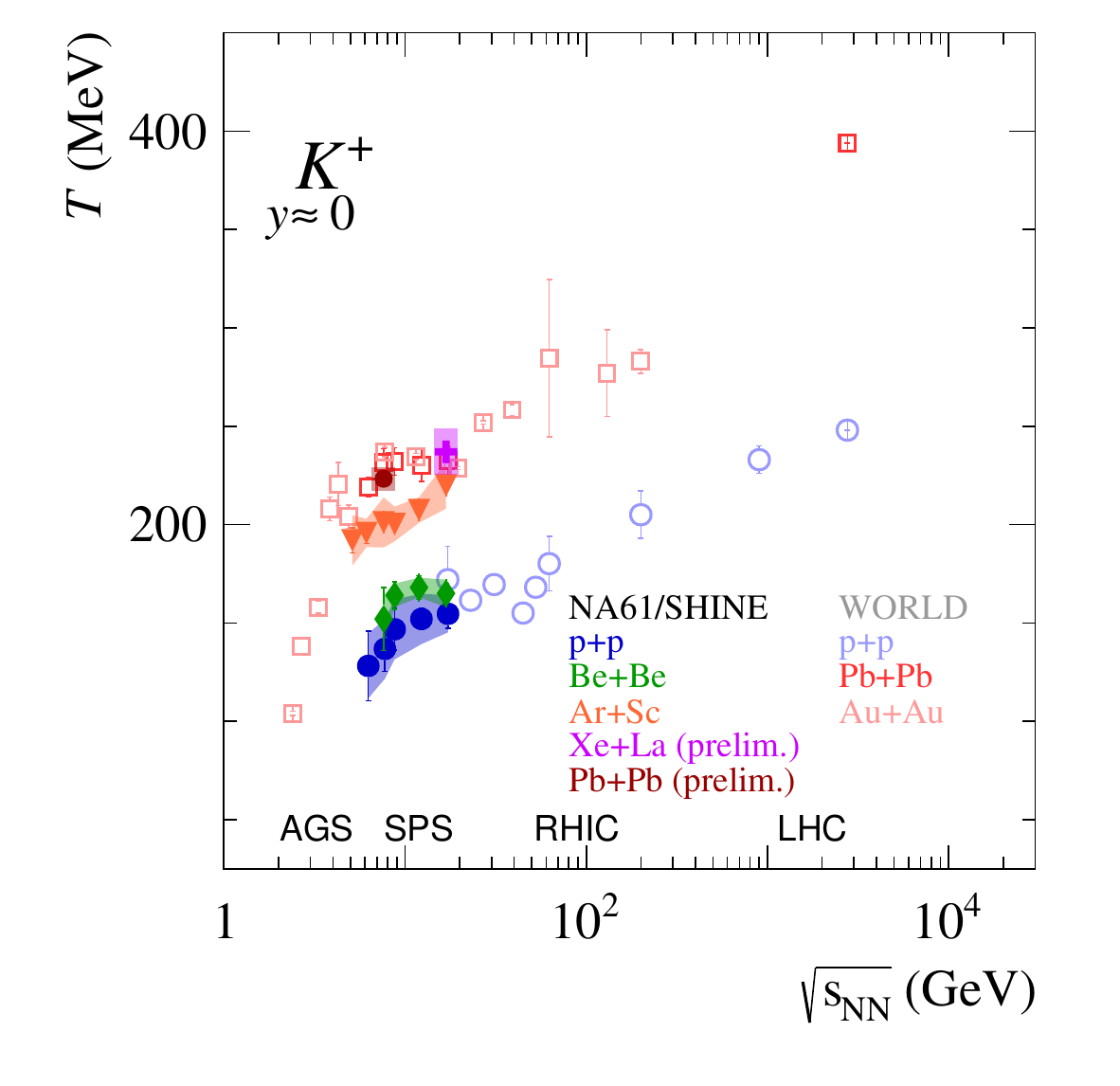}
    \caption{The energy dependence of the $K^+/\pi^+$ ratio in $4\pi$ phase space (\textit{left}) and inverse slope parameter of $p_{\mathrm{T}}$ spectra of $K^+$ (\textit{right}). See Ref.~\cite{NA61SHINE:2023epu} for Ar+Sc results and references to published NA61/SHINE p+p and Be+Be data, as well as world data. For NA61/SHINE points, vertical bars denote statistical uncertainties, and color bands  --  systematic ones.}
    \label{fig:onset}
\end{figure}

Figure \ref{fig:onset} (\textit{left}) presents the $K^+/\pi^+$ ratio as a function of center-of-mass energy $\sqrt{s_{NN}}$. The preliminary result for central Pb+Pb collisions measured at $30A$ $\mathrm{GeV}/c$ $(\sqrt{s_{NN}} \approx 7.6 \, \mathrm{GeV})$, agrees with the NA49 data \cite{NA49:2007stj}. The nonmonotonic behavior in the Pb+Pb data signifies a sharp change in the strangeness production mechanism and is considered as the \textit{horn} signature of the onset of deconfinement.

The inverse slope parameter $T$ was extracted from fits to the $\mathrm{d}^2n/\mathrm{d}y\mathrm{d}p_\mathrm{T}$ spectra of kaons. The NA61/SHINE points were obtained either close to mid-rapidity ($p$+$p$, Be+Be, Ar+Sc) or at $0.4 < y < 0.6$ (Xe+La) and $0.8 < y < 1.0$ (Pb+Pb). The results are presented in Fig.~\ref{fig:onset} (\textit{right}). A plateau in the energy dependence of \textit{T}, known as the \textit{step}, is visible and suggests a softening of the equation of state, as expected during a phase transition involving a mixed phase of hadrons and deconfined matter.

\section*{Summary}
This study presents new results from the NA61/SHINE experiment on central Pb+Pb collisions at a beam momentum of $30A$ $\mathrm{GeV}/c$, focusing on the production of $K^+$, $K^-$, and $\pi^-$ mesons. Particle identification was carried out using $\mathrm{d}E/\mathrm{d}x$ method. Charged kaon $p_T$ and $y$ spectra were fitted using exponential and gaussian functions, respectively, allowing for the extraction of total yields. The $\pi^-$ spectra were obtained via the $h^-$ method, with corrections applied for the contamination based on model predictions. The $K^+/\pi^+$ ratio was calculated by combining measured and extrapolated spectra, with $\pi^+$ multiplicities estimated using the $\pi^+/\pi^-$ ratio from NA49. The NA61/SHINE measurement at $\sqrt{s_{NN}} \approx 7.6 \, \mathrm{GeV}$ agrees with the corresponding NA49 data point. The overall energy dependence of the $K^+/\pi^+$ ratio is consistent with the \textit{horn} structure, which is interpreted as a signature of the onset of deconfinement. Additionally, the inverse slope parameter $T$ shows a plateau at SPS energies, referred to as the \textit{step}, indicating a softening of the equation of state. Together, these observables support the hypothesis that the deconfinement phase transition occurs in this energy region.

\printbibliography

\end{document}